\documentclass[%
 reprint,
 amsmath,amssymb,
 aps,
pra,
floatfix,
]{revtex4-1}
\usepackage{graphicx}
\usepackage{dcolumn}
\usepackage{bm}

\begin{document}

\preprint{APS/123-QED}

\title{Tri-partite non-maximally entangled mixed states as a resource for optimum controlled quantum teleportation fidelity}

\author{K.G. Paulson}\altaffiliation[paulsonkgeorg@gmail.com]{}\affiliation{ Department of Physics, Pondicherry University, Puducherry 605 014, India\\ Department of Physical Sciences, Indian Institute of Science Education and Research Kolkata, Mohanpur-741246, West Bengal, India}
 
\author{Prasanta  K. Panigrahi}\email{pprasanta@iiserkol.ac.in}{}\affiliation{Department of Physical Sciences, Indian Institute of Science Education and Research Kolkata, Mohanpur-741246, West Bengal, India}




\date{\today}

\begin{abstract}
Three-qubit mixed states are used as  a channel for controlled quantum teleportation (CQT) of single-qubit states. The connection between different channel parameters to achieve maximum controlled teleportation fidelity is investigated. We show that for a given multipartite entanglement and mixedness, a class of non-maximally entangled mixed $X$ states ($X-$NMEMS) achieves optimum controlled quantum teleportation fidelity, interestingly  a class of maximally entangled mixed $X$ states ($X-$MEMS) fails to do so. This demonstrates, for a given spectrum and mixedness, that $X-$MEMS are not sufficient to attain optimum controlled quantum teleportation fidelity, which is in contradiction with the traditional quantum teleportation  of single qubits.  In addition, we show that biseparable $X-$NMEMS, for a certain range of mixedness, are useful as a resource to attain high controlled quantum teleportation fidelity, which  essentially lowers the requirements of quantum channels for CQT.

\end{abstract}

\maketitle


\section{ Introduction } \label{sec:level1}
Quantum teleportation is the process of transferring quantum states across two parties separated by large distance without traversing the actual distance between them~\cite{Bennett1993}. In the celebrated teleportation protocol, a single qubit's state is teleported between  two parties, where the  maximally entangled bipartite pure state shared by both parties acts as a quantum channel for the process. Teleportation fidelity determines the  success of quantum teleportation; it is defined as the  overlap of the state to be teleported and the output state at the  receiver's end. It can be considered as  an ascribed characteristic  of  the quantum channel used for the teleportation of an arbitrary quantum state. For a pure quantum channel, the existence of a monotonic relationship between entanglement and teleportation fidelity is well known~\cite{Popescu1994, Horodecki1996a,muralidharan2008}. In reality, quantum systems are open; interaction of the system with surroundings changes the properties  of quantum states in general.  Hence, the exploration of quantum states in noisy environments for implementing various quantum information processing protocols has attracted wide attention. In~\cite{Popescu1994, Horodecki1996a}, it is shown that mixed quantum states can also be used as a channel to achieve imperfect teleportation. In the case of a mixed entangled teleportation channel, there exists no monotonic relationship between entanglement and teleportation fidelity, i.e., a higher value of the entanglement of quantum channel is not sufficient to achieve maximum fidelity~\cite{Paulson2014}.  The connection among different parameters of the quantum channel~\cite{Popescu1994,Horodecki1996a,adhikari2008,Mazzola2010, paulson2016} should be known for the wise usage of channels for quantum teleportation under the effects of noise. For mixed quantum channels,  both mixedness and entanglement contribute to the success of teleportation~\cite{Paulson2014}.\newline
Manipulation of multipartite qubits~\cite{Dur2000,rao2008,de2013, de2017,Shreya2019} is an important task to scale up  the quantum based technology efficiently. A multipartite variant of quantum teleportation  has been proposed in \cite{karlsson1998quantum}, and it is known as controlled quantum teleportation (CQT). In CQT, an arbitrary single-qubit state is  transferred from  sender to receiver only with the permission of the controller. The authority power  of the  controller to decide the success or failure of teleportation for tri-partite CQT protocol shows its   difference from the bipartite one. Recently, Barasinski et.al., experimentally implemented controlled quantum teleportation of single-qubit state on linear optical devices~\cite{barasinski2019Exp} and discussed the possibilities of controlled quantum teleportation by lowering the requirements of quantum channels.\newline
Conditioned and nonconditioned fidelity are two  quantities that  are measured with and without the permission of the  controller, characterizing the CQT protocol. It is assumed that in CQT, $F_{CQT}$ (conditioned fidelity) should be always greater than the classical limit, whereas the value of $F_{NC}$  (nonconditioned fidelity)~\cite{karlsson1998quantum,li2014,jeong2016} cannot exceed the classical limit $\frac{2}{3}$ $(F_{NC}\leq\frac{2}{3})$. The classical limit of  nonconditioned fidelity is calculated for the set of pure input states that are  chosen according to the Haar measure~\cite{Massar1995}.  The control power (CP), a quantity to define the authority of the controller in CQT, is estimated as the difference of conditioned and nonconditioned fidelity.\newline
As is known, for bipartite quantum states, purity of the quantum channel along with entanglement~\cite{Bose2000,Paulson2014,Paulson2017} play  a significant role in the implementation of quantum teleportation process with maximum achievable fidelity.   Different classes of states are considered as quantum channels for teleportation, among which a class of $X$ states, having non zero diagonal and antidiagonal elements, deserves  special attention~\cite{hagley1997,pratt2004,wang2006}. In the case of the bipartite qubit system, a given density matrix can be  unitarily  transformed to $X$ state with same  degree of entanglement and spectrum ~\cite{Hedemann2013, mendoncca2014}. Thus quantum states in $X$ structure form an important class of density matrices in general and are used as a representative class of states for quantum information processing.\newline
The use of tripartite quantum states as a channel for controlled quantum teleportation and the estimation of controlled teleportation fidelity  for  $X$ states are shown in~\cite{barasinski2018}. Both maximally  and non-maximally entangled pure Greenberger-Horne-Zeilinger (GHZ) like states act as quantum channels for CQT. In ~\cite{barasinski2019} it is shown,  how genuine multipartite entanglement (GME) and CP affect the controlled quantum teleportation  fidelity for a class of $X$ states. Purity of tripartite  quantum states is an important  parameter that affects quantum correlations and investigation of the efficacy of tripartite quantum states for CQT will not be conclusive without accounting for the purity of quantum channel along with other channel parameters. We fill this gap by a detailed investigation on the performance of mixed quantum channel for controlled quantum teleportation.\newline
We systematically investigate the roles played by various parameters, like purity, entanglement and control power of tripartite qubit states in achieving optimum controlled quantum teleportation fidelity ($F_{CQT}$). For this purpose, we consider different classes of multipartite $X$ states and analyze their performance as  CQT channels. First, we examine the faithfulness of a class of rank dependent  maximally entangled mixed $X$ states ($X$-MEMS), defined for a given spectrum of eigenvalues and linear entropy as a CQT resource. Since the performance of $X$-MEMS as a CQT  channel is not optimum, a  class of tripartite non-maximally entangled mixed $X$ states ($X$-NMEMS) is constructed and its teleportation fidelity is estimated. We show that  our  class of $X$-NMEMS outperforms $X$-MEMS as a quantum channel for CQT and rank-2 $X$-NMEMS gives maximum achievable teleportation fidelity for a given entanglement and mixedness as shown in~\cite{barasinski2019}. This clearly demonstrates that  CQT protocol lowers the requirements of the quantum channel for the successful quantum teleportation of a single qubit's state. At high value of mixedness, $X$-NMEMS become biseparable. Even with the biseparability condition, $X$-NMEMS are found to give high values for controlled quantum teleportation fidelity above the classical limit. This high value of  fidelity of the biseparable quantum channel  is a direct evidence  that  mixed tripartite quantum states can lower the requirements of the quantum channel for  successful controlled teleportation.\newline
From our investigation on tripartite mixed quantum channels, we show that tri-partite $X-$MEMS are not sufficient  to achieve optimum CQT fidelity, whereas optimum controlled quantum teleportation fidelity is achieved using a class of $X-$NMEMS. Even though genuine multipartite entanglement of  $X-$NMEMS vanishes for high values of mixedness, the process of  controlled quantum teleportation of single qubit state is enabled by the  biseparability nature of $X-$NMEMS. These results, which lower the requirements of quantum channel are quite important for the  experimental realization of controlled quantum teleportation in noisy environment.\newline 
The present paper is organized as follows. In  Sec.~\ref{sec1}, we discuss the prerequisites for implementing the CQT protocol.  Section.~\ref{sec2} contains two subsections, first subsection deals with the construction of tripartite qubit $X$-MEMS, its usefulness for controlled quantum teleportation. It is  followed by the construction of a  class of $X$-NMEMS and its  efficacy as a quantum channel for CQT is analyzed in the second subsection. Results and discussion in Sec.~\ref{sec6} are followed by the concluding section (Sec.~\ref{con}).

\section{Preliminaries}\label{sec1}
Below, we define different parameters GME, teleportation fidelity, control power and linear entropy,  which  characterize the tripartite mixed entangled quantum channels for controlled quantum teleportation.
\subsection{Genuine Multipartite Entanglement(GME)}
The three-qubit symmetric mixed  $X-$states~\cite{yu2007} are defined with diagonal elements denoted by 
$a_{1},a_{2},a_{3},a_{4},b_{1},b_{2},b_{3},b_{4}\geq1$ and 
antidiagonal elements given by
$z_{1},z_{2},z_{3},z_{4},z_{1}^{*},z_{2}^{*},z_{3}^{*},z_{4}^{*}$.
The genuine multipartite entanglement (GME) of a three-qubit $X$ state  is given as,
\begin{equation}
    C_{GME}=2\,max\{0,\vert z_{j}\vert-w_{j}\},
\end{equation}
where $\sum_{i}(a_{i}+b_{i})=1$ and $w_{j}=\sum_{k\neq j}\sqrt{a_k b_k}$. The positivity criterion of the $X$-matrix 
is satisfied with the condition $\vert z_{i}\vert\leq\sqrt {a_i b_i}$. The tripartite $X$ states are entangled for $0< C_{GME}\leq1$ and $C_{GME}$ is zero for biseparable states~\cite{rafsanjani2012}. 
\subsection{Controlled quantum teleportation fidelity}
Here, we describe the protocol of controlled quantum teleportation of a single qubit's state via the tripartite qubit channel. Consider that  three parties, labeled as $A$, $B$ and $C$,  shared an entangled three-qubit quantum state $\rho_{abc}$, which  acts as a channel connecting them to each other. Suppose party $A$ wants to teleport an unknown state of qubit $d$ to $B$ with the consent of party $C$. At this moment  controller $C$ makes an orthogonal measurement on his qubit $c$, with $\zeta$ as the measurement outcome. This results in the projection of entangled channel $\rho_{abc}$ onto the two-qubit state~\cite{jeong2016} $\rho^{\zeta}_{ab}$;
\begin{equation}
    \rho^{\zeta}_{ab}=\frac{Tr_{c}[1_{2}\otimes 1_{2}\otimes\vert\zeta\rangle\langle\zeta\vert U\rho_{abc}1_{2}\otimes 1_{2}\otimes U^{\dag}\vert\zeta\rangle\langle \zeta\vert]}{\langle \zeta\vert U\rho_{c}U^\dag\vert\zeta\rangle}.
\end{equation}
Here  $1_{2}$,  a $2\times2$ identity matrix, acts on the qubit's state with observers $A$ and $B$, $U$  a  $2\times2$,   unitary matrix along with projection operation act on the qubit's state with observer $C$ and $\rho_c=Tr_{ab}[\rho_{abc}]$. Following this, party $A$ makes a joint orthogonal measurement on qubits $a$ and $d$ and communicates the results to $B$, and appropriate unitary operations on qubit $b$ completes the process of CQT. The controlled quantum teleportation fidelity $F_{CQT}(\rho)$ in this scenario is defined as,
\begin{equation}\label{fcqt}
    F_{CQT}(\rho)=\frac{2 max_{U}[\sum_{\zeta=0}^{1}\langle \zeta\vert U\rho_{c}U^{\dag}\vert\zeta\rangle f(\rho^{\zeta}_{ab})]+1}{3}.
\end{equation}
We have nonconditioned teleportation fidelity (without the controllers participation) given as,
\begin{equation}\label{fnc}
    F_{NC}(\rho)=\frac{2 f(\rho_{ab})+1}{3},
\end{equation}
where $f(\rho)$  is the fully entangled fraction~\cite{Bennett1996a, Horodecki1999, ma2011} and $\langle\zeta\vert U\rho U^{\dag}\vert\zeta\rangle$ is the maximum probability of receiving outcome $\zeta$. The fidelities derived in Eqs.~(\ref{fcqt}) and (\ref{fnc}) are estimated for  general three-qubit mixed $X$ states ~\cite{barasinski2018} as follows,
\begin{equation}
    F_{CQT}(\rho_{X})=max\{ F_{CQT}^{1},F_{CQT}^{2},F_{CQT}^{3},F_{CQT}^{4}\}
\end{equation}
where,
\begin{eqnarray}
\nonumber
F_{CQT}^{1}=\frac{3+\vert\Delta_{1}\vert+4(\vert z_{1}\vert+\vert z_{4}\vert)}{6}\\ \nonumber
F_{CQT}^{2}=\frac{3+\vert\Delta_{1}\vert+4(\vert z_{2}\vert+\vert z_{3}\vert)}{6}\\ \nonumber
F_{CQT}^{3}=\frac{3+\sqrt{\Delta_{2}^2+16(\vert z_{1}\vert+\vert z_{4}\vert)^2}}{6}\\ 
F_{CQT}^{4}=\frac{3+\sqrt{\Delta_{2}^2+16(\vert z_{2}\vert+\vert z_{3}\vert)^2}}{6}.\\ \nonumber
\end{eqnarray}
Here $\Delta_{1}=a_{1}-a_{2}-a_{3}+a_{4}+b_{1}-b_{2}-b_{3}+b_{4}$,  $\Delta_{2}=a_{1}-a_{2}+a_{3}-a_{4}-b_{1}+b_{2}-b_{3}+b_{4}$. The non-conditioned teleportation fidelity of the state $\rho_{X}$ is,
\begin{equation}
    F_{NC}(\rho_{X})=\frac{3+\vert\Delta_{1}\vert}{6}.
\end{equation}
The influence of the control qubit in CQT process is quantified by estimating CP and is defined as,
\begin{equation}
    CP(\rho_X)=F_{CQT}(\rho_X)-F_{NC}(\rho_X).
\end{equation}
The two conditions, $F_{CQT}(\rho)>\frac{2}{3}$ and $F_{NC}(\rho)\leq\frac{2}{3}$ should be satisfied by tripartite quantum channels to ensure the active participation of the controller in the controlled quantum teleportation process. 
Mixedness of quantum states is an important parameter that influences fidelity of controlled quantum teleportation. We use linear entropy to estimate the mixedness of a state, which is defined for  a multipartite qubit state $\rho$  as,
\begin{equation}
    S_{L}(\rho)= \frac{2^N-1}{2^N}[1-Tr(\rho^2)].
\end{equation}.
Here $N$ is the number of qubits and $Tr(\rho^2)$ is the purity of the multipartite quantum state. Mixed states satisfy the condition  $0< S_{L}(\rho)\leq1$  and $S_{L}(\rho)=0$ for pure states.
\section{Mixed $X$ states, a resource for controlled quantum teleportation}\label{sec2}
In this section, we investigate  in detail the  mixed three-qubit $X$ states as a resource  for controlled  quantum teleportation. We show how  purity  and other quantum correlations of tripartite qubit states are connected to each other for their usage as a CQT channel. From the study  of the bipartite mixed quantum channel as a resource for  teleportation of   single-qubit states, we infer the non-trivial dependence of teleportation fidelity on mixedness and entanglement of the quantum channel. In \cite{Paulson2014, Paulson2017}, one of the present  authors has shown the existence of rank dependent bounds on mixedness and entanglement of quantum states for their usefulness for successful quantum teleportation. Among bipartite qubit quantum channels, a class of MEMS~\cite{Ishizaka2000,Verstraete2001,Munro2001,Wei2003} gives maximum teleportation fidelity for a given mixedness and entanglement. This demonstrates its  importance in  investigating the efficacy of  mixed entangled teleportation channels in higher-dimensional state space. We address this situation by considering tripartite mixed $X$ quantum channel for CQT.
\subsection{Tri-partite  maximally entangled mixed $X$ states }\label{sec3}
\begin{figure}[!htb]
    \centering
    \includegraphics[height=65mm,width=1\columnwidth]{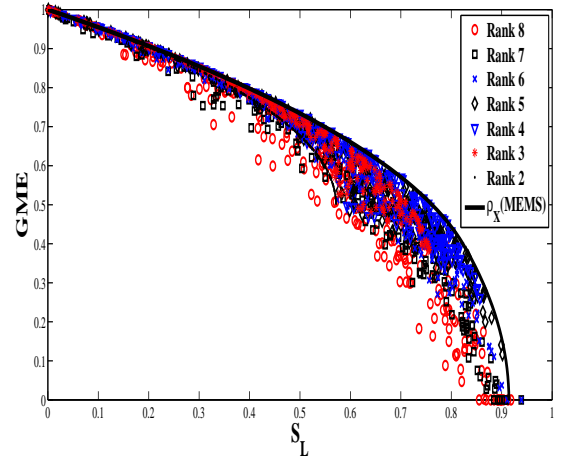}
    \caption{Genuine multipartite entanglement of various rank $X$-MEMS  is plotted as a function of linear entropy. Ranks of the states vary from $2$ to $8$. The entanglement of  three-qubit $X$ MEMS (Eq.~(\ref{xmemssl})) defined with respect to purity of the quantum states possesses maximum value for the range of values of linear entropy.}
    \label{slvsgme_mems}
\end{figure}

\begin{figure}[!htb]
    \centering
    \includegraphics[height=65mm,width=1\columnwidth]{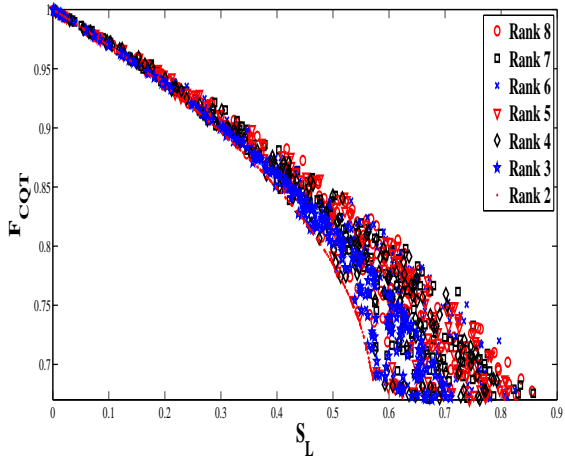}
    \caption{ Controlled teleportation fidelity $F_{CQT}$ of $X$-MEMS is given as a function of linear entropy $S_{L}$. It is clear that higher rank MEMS gives higher value of teleportation fidelity for a fixed linear entropy.}
    \label{slvsf_mems}
\end{figure}

\begin{figure}[!htb]
    \centering
\includegraphics[height=65mm,width=1\columnwidth]{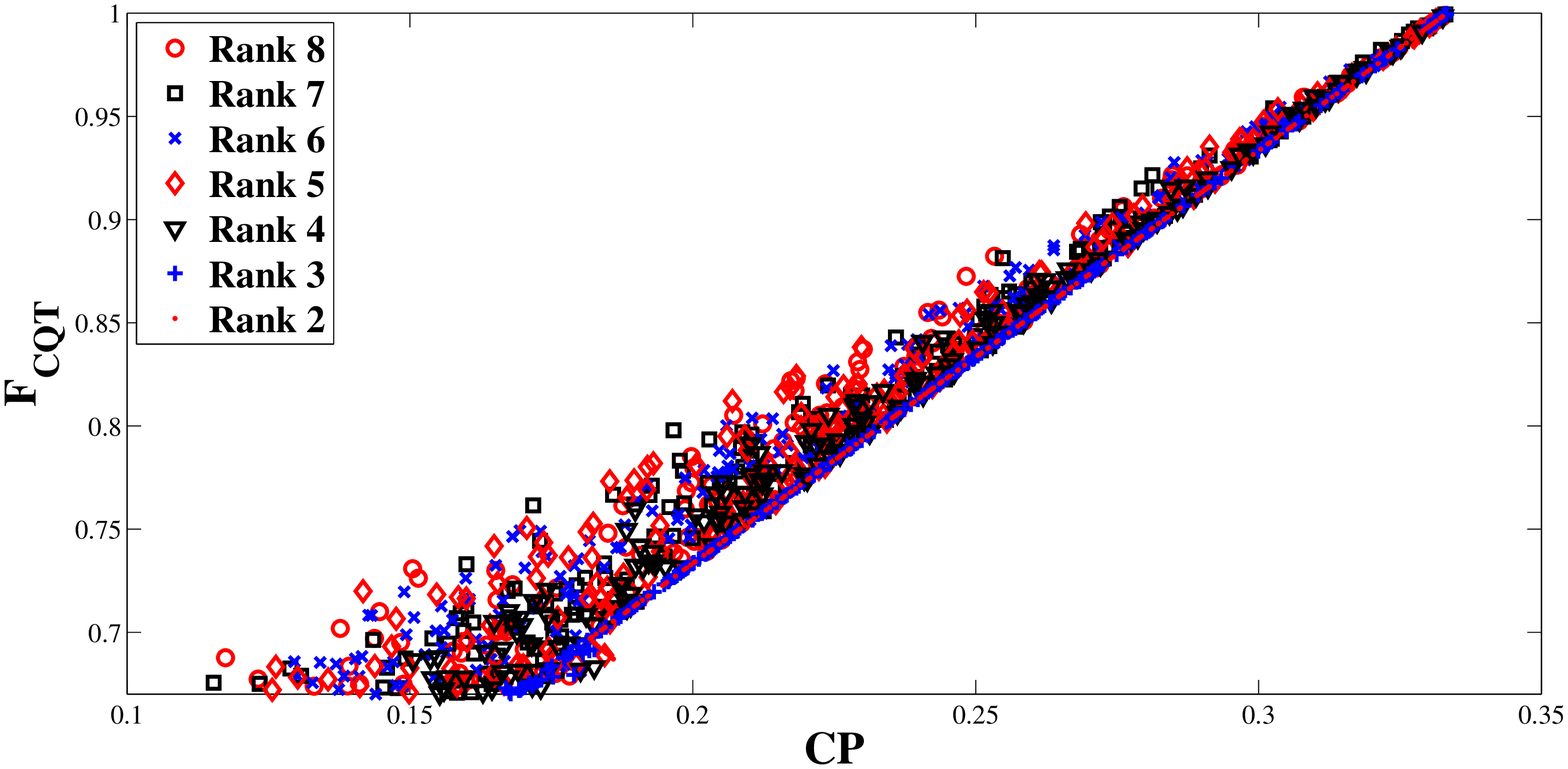}
    \caption{$X$-MEMS of all ranks are generated and the controlled teleportation fidelity $F_{CQT}$ of $X$-MEMS is plotted as a function of control power CP.}
    \label{cpvsf_mems}
\end{figure}

The genuine maximally entangled mixed $X$ states for $N$-qubits are given in~\cite{mendoncca2015} for a given spectrum of eigenvalues. The class of three-qubits $X$-MEMS as  a convex sum of maximally entangled pure GHZ and separable states is given as,
\begin{eqnarray}
\nonumber
\rho(X)_{MEMS}=p_{1}\vert GHZ_{1}^{+}\rangle\langle GHZ_{1}^{+}\vert+ p_{2}\vert 001\rangle\langle 001\vert\\ \nonumber
+p_{3}\vert 010\rangle\langle 010\vert+p_{4}\vert011\rangle\langle 011\vert+p_{5}\vert GHZ_{1}^{-}\rangle\langle GHZ_{1}^{-}\vert\\ \nonumber
+p_{6}\vert 100\rangle\langle 100\vert+ p_{7}\vert 101\rangle\langle 101\vert+ p_{8}\vert 110\rangle\langle 110\vert.  \\
\end{eqnarray}
Where $p_{1}\geq p_{2} \geq p_{3} \geq p_{4} \geq p_{5} \geq p_{6} \geq p_{7} \geq p_{8}\geq 0$ are the eigenvalues of density matrix $\rho(X)_{MEMS}$ and $\ p_{1}+ p_{2}+p_{3}+p_{4}+p_{5}+p_{6}+p_{7}+p_{8}=1 $, satisfies the normalization condition of the density matrix. The maximally entangled  three-qubit  GHZ state basis is given as,
\begin{eqnarray}\label{ghz_basis}
\nonumber
\vert GHZ_{1}^{\pm}\rangle=\frac{1}{\sqrt{2}}[\vert 000\rangle\pm\vert 111\rangle]\\ \nonumber
\vert GHZ_{2}^{\pm}\rangle=\frac{1}{\sqrt{2}}[\vert 001\rangle\pm\vert 110\rangle]\\ \nonumber
\vert GHZ_{3}^{\pm}\rangle=\frac{1}{\sqrt{2}}[\vert 010\rangle\pm\vert 101\rangle]\\
\vert GHZ_{4}^{\pm}\rangle=\frac{1}{\sqrt{2}}[\vert 011\rangle\pm\vert 100\rangle].
\end{eqnarray}
It is shown that the given density matrix $\rho(X)_{MEMS}$ possesses maximum value of GME for a given spectrum of eigenvalues $\{\Lambda\}$. We calculate the  GME of $\rho(X)_{MEMS}$  and it is given by,
\begin{equation}\label{Gmems}
    C^{*}(\rho(X))=max\{0,p_{1}-p_{5}-2[\sqrt{p_{2}p_{8}}+\sqrt{p_{3}p_{7}}+\sqrt{p_{4}p_{6}}]\}.
\end{equation}
If GME of a given $\rho(X)$  is equal to $C^{*}(\rho(X))$, then the state $\rho(X)$ belongs to the class of $\rho(X)_{MEMS}$. \newline
The maximally entangled mixed three-qubit $X$ states, defined with respect to the  mixedness of quantum states~\cite{agarwal2013} are given as,
\begin{equation}\label{xmemssl}
\rho(X)=\left(
\begin{array}{cccccccc}
f(\gamma)& 0 & 0 & 0 & 0 & 0 & 0 &  \gamma\\
0& g(\gamma) & 0 & 0 & 0 & 0 & 0 & 0\\
0& 0 & g(\gamma) & 0 & 0 & 0 & 0 & 0\\
0 & 0 & 0 & g(\gamma) & 0 & 0 & 0 & 0\\
0& 0 & 0 & 0 & 0 & 0 & 0 & 0\\
0& 0 & 0 & 0 & 0 & 0 & 0 & 0\\
0& 0 & 0 & 0 & 0 & 0 & 0 & 0\\
\gamma& 0 & 0 & 0 & 0 & 0 & 0 & f(\gamma)
\end{array}\right)
\end{equation}
where,
\begin{equation}
f(\gamma)=
\begin{cases}
1/5, &  0\leq\gamma\leq 1/5\\
            \gamma,& 1/(5)< \gamma\leq 1/2
            \end{cases}
\end{equation}
and
\begin{equation}
    g(\gamma)=
    \begin{cases}
    1/5,& 0\leq\gamma \leq1/5  \\
    (1-2\gamma)/3,& 1/5<\gamma\leq1/2.
    \end{cases}
\end{equation}
The tri-partite $X-$MEMS, defined with respect to purity, is of rank $4$ and $5$. The GME of the above defined maximally entangled mixed state is $max[0, 2 \vert \gamma\vert]$. The GME of three-qubit $X-$MEMS of different ranks as a function of linear entropy is given in  Fig.\ref{slvsgme_mems}.
From the Fig.\ref{slvsgme_mems}, it is clear that, rank $4$ and $5$ $X$-MEMS possess the highest value of entanglement for a fixed linear entropy. The tripartite $X$-MEMS, defined with respect to purity,  possesses maximum achievable multipartite entanglement among all rank dependent $X$-MEMS. Here, we use this  class of tripartite qubit $X$-MEMS  as a channel for controlled quantum teleportation and show how the teleportation fidelity of different rank  MEMS varies as a function of mixedness  and other quantum correlations. The controlled quantum teleportation fidelity of $X-$MEMS is given as,
\begin{eqnarray}\label{fxmems}
\nonumber
    F_{CQT}(MEMS)=\frac{1}{6}(3+2(p_{1}-p_{5})\\
    +\vert p_{1}-p_{2}-p_{3}+p_{4}+p_{5}+p_{6}-p_{7}-p_{8}\vert).
\end{eqnarray}
The nonconditioned fidelity takes the value $\frac{1}{6}(3+\vert p_{1}-p_{2}-p_{3}+p_{4}+p_{5}+p_{6}-p_{7}-p_{8}\vert)$ and is always less than or equal to the classical limit of fidelity $\frac{2}{3}$. The CQT fidelity of $X-$MEMS as a function of linear entropy is given in Fig.\ref{slvsf_mems}.
From  Fig.\ref{slvsf_mems}, in which teleportation fidelity of $X$-MEMS of ranks, varying from $2$ to $8$ is analyzed as a function of linear entropy, we infer that higher rank maximally entangled mixed states survive as a CQT channel for higher value of  mixedness.
\begin{figure}[!htb]
    \centering
    \includegraphics[height=65mm,width=1\columnwidth]{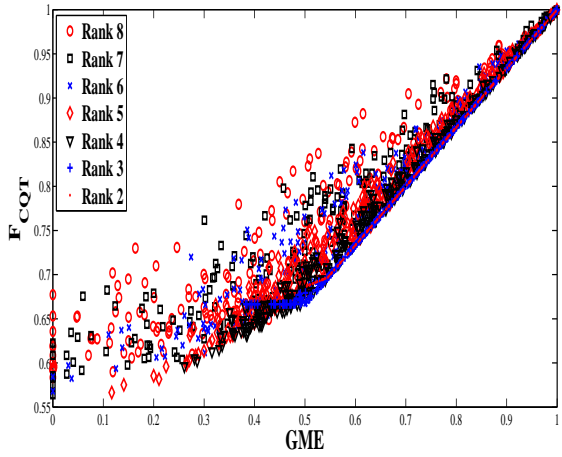}
    \caption{Controlled quantum teleportation fidelity $F_{CQT}$ of rank dependent tri-partite $X$-MEMS is plotted as a function of genuine multipartite entanglement, it is shown that higher rank states posses higher values of teleportation fidelity for lower values of GME.}
    \label{gmevsf_mems}
\end{figure}
In Fig.\ref{gmevsf_mems}, we analyze the controlled teleportation fidelity of different rank $X$-MEMS as a function of genuine multipartite entanglement. It is seen that higher rank states possess higher value of teleportation fidelity for lower values of GME instead of maximally entangled mixed  $X$ states (with maximum GME) defined with respect to purity.This implies that there exists no monotonic relationship between  entanglement and teleportation fidelity in the case of tripartite mixed channels.\newline
The control parameter is another quantity that captures the authority of the controller's qubit in the process of  CQT. Control quantum teleportation fidelity as a function of control power for different rank $X$-MEMS is given in  Fig.\ref{cpvsf_mems}.
The CQT fidelity of different rank $X$-MEMS under the authority of the controller qubit holds the bounds proposed in~\cite{barasinski2019}.
The boundaries for maximally entangled mixed $X$ states of rank $r$ $(2\leq r\leq 8)$ are constructed, by identifying the spectrum of eigenvalues as $p_{1}=\frac{1+(r-1)p}{r}$ and rest of the $r-1$ eigenvalues equal to $\frac{1-p}{r}$. These  boundary states act as an upper bound of corresponding rank dependent $X-$MEMS for the curves in which teleportation fidelity is analyzed as a function of linear entropy and multipartite entanglement. Since the CQT fidelity of $X-$MEMS is not optimum, we construct a  class of non maximally entangled mixed states, $X-$NMEMS. The details of the investigation and its performance as a quantum channel for controlled quantum teleportation  are discussed in the next section. 

\subsection{ Tri-partite non-maximally entangled mixed $X$ states}\label{sec5}
\begin{figure}[!htb]
    \centering
\includegraphics[height=65mm,width=1\columnwidth]{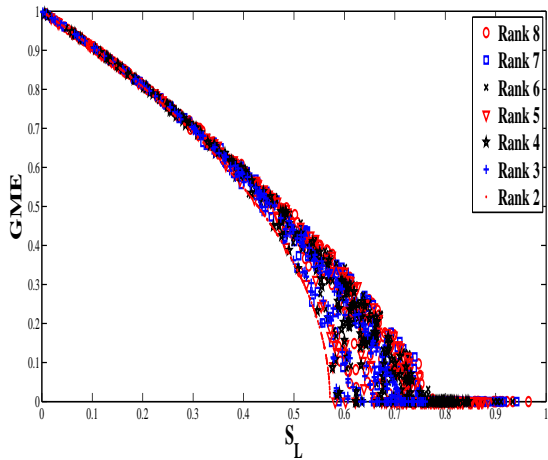}
    \caption{Genuine multipartite entanglement  of tripartite $X-$NMEMS of all ranks is plotted as a function of linear entropy $S_{L}$.}
    \label{slvsgme_nmems}
\end{figure}

\begin{figure}[!htb]
    \centering
\includegraphics[height=65mm,width=1\columnwidth]{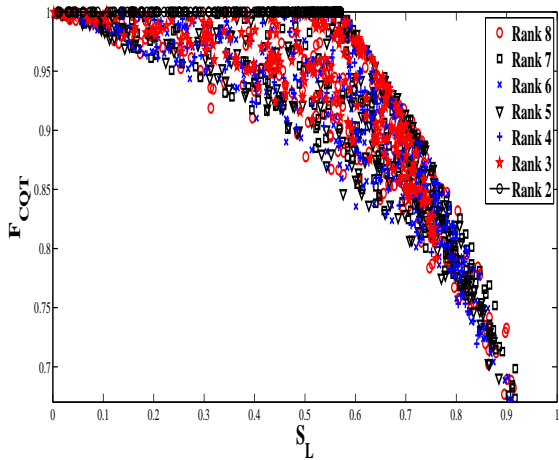}
    \caption{Tri-partite controlled quantum teleportation fidelity $(F_{CQT})$ of $X-$NMEMS of various ranks is plotted for a given value of linear entropy $(S_{L})$. Rank $2$  $X-$NMEMS gives the maximum achievable CQT fidelity.}
    \label{slvsf_nmems}
\end{figure}

In this section, we construct  a class of tripartite  $X$-NMEMS; their performance as a CQT channel is investigated and is compared  with  that of $X$-MEMS. We show that our new class of $X$-NMEMS is a potential candidate for  controlled quantum teleportation  of a qubit's state through three-qubit quantum channel at a high value mixedness and a low value of entanglement. The class of $X$-NMEMS, as a convex combination of maximally entangled GHZ states in Eq.~(\ref{ghz_basis}), is given as
\begin{eqnarray}
\nonumber
\rho(X)_{NMEMS}=p_{1}\vert GHZ_{1}^{+}\rangle\langle +GHZ_{1}^{+}\vert+p_{2}\vert GHZ_{4}^{+}\rangle\langle GHZ_{4}^{+}\vert\\ \nonumber
+p_{3}\vert GHZ_{2}^{+}\rangle\langle GHZ_{2}^{+}\vert +p_{4}\vert GHZ_{3}^{+}\rangle\langle GHZ_{3}^{+}\vert\\ \nonumber
+p_{5}\vert GHZ_{1}^{-}\rangle\langle GHZ_{1}^{-}\vert +p_{6}\vert GHZ_{4}^{-}\rangle\langle GHZ_{4}^{-}\vert\\ \nonumber
+p_{7}\vert GHZ_{2}^{-}\rangle\langle GHZ_{2}^{-}\vert+p_{8}\vert GHZ_{3}^{-}\rangle\langle GHZ_{3}^{-}\vert .  \\
\end{eqnarray}
The eigenvalues $p_{i}^{'s}$ of non-maximally entangled mixed $X$ states satisfy the  conditions of normalization and positivity  discussed in  Sec.~\ref{sec3}. We investigate the details of the $X$-NMEMS quantum channel for CQT and show that tri-partite mixed entangled states lower the requirements of the controlled quantum teleportation channel.\newline
The genuine multipartite entanglement of the  non-maximally entangled mixed state is estimated as,
\begin{equation}\label{cxnmems}
    C(\rho(X)_{NMEMS})=(p_{1}-p_{5})-(p_{2}+p_{3}+p_{4}+p_{6}+p_{7}+p_{8}).
\end{equation}
The above-constructed tri-partite $X$ state does not fall in the class of $X-$MEMS, since the estimated GME of $X$ states is not equal to $C^{*}\rho(X)$ in Eq.~(\ref{Gmems}).
The calculated   controlled teleportation fidelity of $X$-NMEMS is given by, 
\begin{eqnarray}\label{fxnmems}
\nonumber
    F_{CQT}({NMEMS})=\frac{1}{6}(3+3(p_{1}+p_{2})\\
              -(p_{3}+p_{4}+p_{5}+p_{6}+p_{7}+p_{8})).
\end{eqnarray}
The classical limit of teleportation fidelity of $X-$NMEMS is, $F_{NC}=\frac{1}{6}(3+\vert p_{1}+p_{2}-p_{3}-p_{4}+p_{5}+p_{6}-p_{7}-p_{8}\vert)$.
The genuine multipartite entanglement of  different rank $X$-NMEMS is plotted as a function of linear entropy in Fig.\ref{slvsgme_nmems}. 
From Fig.\ref{slvsgme_nmems}, it is clear that the new class of tripartite $X-$NMEMS possesses a lower value of GME as compared to that of  $X-$MEMS for a defined  spectrum of eigenvalues and linear entropy. Moreover, from  Fig.\ref{slvsgme_nmems}, we infer that the entanglement of three-qubit $X-$NMEMS increases as rank increases, which is not the case for $X-$MEMS. We use this non-maximally entangled mixed state as a CQT channel and teleportation fidelity as a function of linear entropy is given in Fig.~\ref{slvsf_nmems}.

\begin{figure}[htb]
    \centering
\includegraphics[height=65mm,width=1\columnwidth]{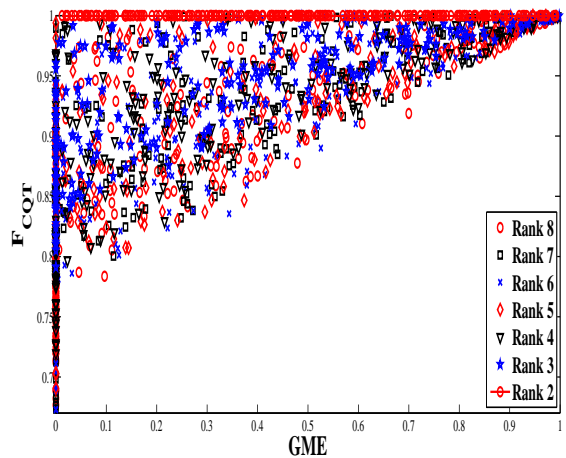}
    \caption{Controlled  quantum teleportation fidelity $F_{CQT}$ of three-qubit $X-$NMEMS is given  as a function of genuine multipartite entanglement  and it is shown that the maximum achievable CQT fidelity is attained for rank-2 $X-$NMEMS for the full range of values of GME.}
    \label{gmesvsf_nmems}
\end{figure}

\begin{figure}[htb]
\centering
\includegraphics[height=65mm,width=1\columnwidth]{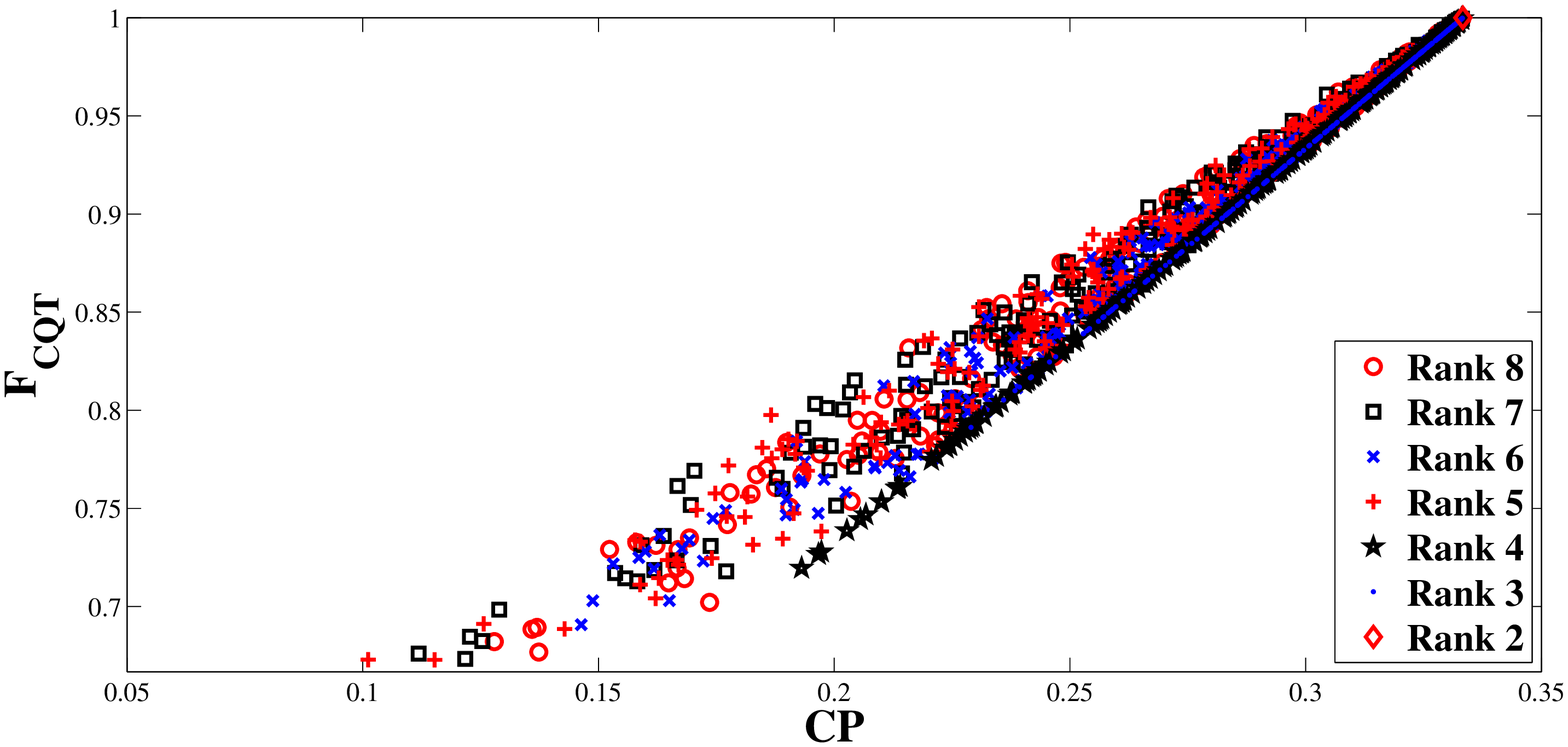}
\caption{Controlled quantum teleportation fidelity $F_{CQT}$ of non-maximally entangled mixed $X$ states ($X-$NMEMS) is plotted as a function of control power CP.}\label{cpvsf_nmems}
\end{figure}

We analyze the performance of our $X$-NMEMS  for the CQT process  as a function of both entanglement and mixedness. It is clear from Eq.~(\ref{fxnmems}) that rank-2 $X-$NMEMS gives maximum achievable controlled teleportation fidelity for all values of entanglement and mixedness. This is in contradiction with the case of the conventional quantum teleportation process. From Fig.~\ref{slvsf_nmems} and Fig.~\ref{gmesvsf_nmems}, wherein controlled teleportation fidelity is analyzed as a function of mixedness and entanglement respectively, it is seen that rank dependent $X-$NMEMS give maximum CQT fidelity in both cases as compared to $X-$MEMS. At the same time, as it is known from Fig.~\ref{slvsgme_nmems}, even though GME of $X-$NMEMS is lower than that of $X-$MEMS for a given purity, its performance as a CQT channel is optimum. This indicates that maximum value of entanglement is not a necessary and sufficient condition to achieve optimum  controlled quantum teleportation fidelity, which is not the case for bipartite quantum channels. At lowest rank (rank $2$), $X-$NMEMS are the  same as the states in \cite{barasinski2018} and they give maximum achievable controlled teleportation fidelity among all mixed states.
The CQT fidelity of $X-$NMEMS is given as a function of control power in Fig.~\ref{cpvsf_nmems}. $X-$NMEMS hold the lower and upper bounds defined for CQT fidelity, in terms of control power and multipartite entanglement.
As we have discussed for maximally entangled mixed multipartite states, the rank dependent boundary $X-$NMEMS are constructed by considering the eigenvalues, $p_{1}=\frac{1+(r-1)p}{r}$ and rest of the $r-1$ eigenvalues equal to $\frac{(1-p)}{r}$. In the case of  $X-$NMEMS, rank dependent boundary states act as lower bounds of respective rank $X-$NMEMS for CQT fidelity, given as a function of multipartite entanglement and mixedness.
\section{Results and Discussions}\label{sec6}
In this paper, we systematically investigated the efficacy of the tripartite mixed entangled state as a resource for controlled quantum teleportation. Mixedness and entanglement jointly decide the efficiency of mixed quantum channels for CQT. To investigate the interdependence of multipartite entanglement, mixedness and control power of the quantum states on the success of controlled quantum teleportation  in detail, we used  a class of tri-partite maximally entangled mixed $X$ states   as a  channel for CQT. The rank dependent performance of $X-$MEMS as a CQT channel has been analyzed as a function of the aforementioned channel parameters and it is shown that the $X-$MEMS do not give optimum controlled quantum teleportation fidelity, as  is true for the bipartite quantum states.\newline
The problem of controlled quantum teleportation via non-maximally entangled pure state has already been studied. Here we extended  this work to the usage of  non-maximally entangled mixed states as a CQT resource. We constructed a  class of non-maximally entangled mixed states and investigated its application as a  controlled  quantum teleportation channel. We showed that a  class of $X-$NMEMS outperforms $X-$MEMS as a CQT channel, for a given mixedness and entanglement. This essentially proves that maximum multipartite entanglement is not sufficient for achieving optimum teleportation fidelity. From  Fig.~\ref{slvsgme_nmems}, it is known that for some  values of linear entropy, the genuine multipartite entanglement of  tripartite $X-$NMEMS becomes zero. Zero GME implies that states are biseparable. From our investigation on tripartite $X-$NMEMS for CQT, it is evident that biseparable states are useful for CQT at a high  degree of mixedness. This result is an important one for the experimental realization of CQT in a noisy environment. For example consider the case of boundary $X-$NMEMS ($\rho_{3}(X)$) of rank $3$: the eigenvalues of $\rho_3(X)$ are $p_{1}=\frac{1+2p}{3}$ and $p_{2}=p_{3}=\frac{1-p}{3}$. We calculated the channel  parameters of $\rho_{3}(X)$ as, $F_{CQT}=\frac{7+2p}{9}$, $F_{NC}=\frac{5+p}{9}$, $GME=max\{0,\frac{4p-1}{3}\}$ and $S_{L}=\frac{16(1-p^2)}{21}$. Multipartite entanglement of $\rho_{3}(X)$ is zero for $0\leq p\leq\frac{1}{4}$; i.e., for high value of mixedness  $S_{L}\geq\frac{15}{21}$, rank-3 boundary $X-$NMEMS possess no genuine multipartite entanglement. However the controlled teleportation fidelity of rank $3$ $X-$NMEMS does not vanish above this value of mixedness, $F_{CQT}=\{\frac{7}{9},\frac{5}{6}\}$ for $\frac{15}{21}\leq S_{L}\leq\frac{16}{21}$. Even for  the biseparability nature of boundary $X-$NMEMS at high values of mixedness, the controlled quantum teleportation fidelity  possesses a high value above the classical limit. \newline
\section{Conclusions}\label{con}
Analysis of the performance of tri-partite rank dependent $X$ states as a resource for controlled quantum teleportation revealed many intriguing properties of multipartite systems that can be exploited for the efficient implementation of  quantum information processing protocols. We showed that for a given multipartite entanglement and mixedness, a  class of non-maximally entangled mixed $X$ states  achieve optimum controlled quantum teleportation fidelity. At the same time, investigation on $X-$MEMS as a resource for CQT proved that tri-partite maximally entangled mixed states fail to attain optimum teleportation fidelity. From our investigation on $X-$NMEMS, we showed that the class of biseparable $X-$NMEMS can also be considered as a potential candidate for CQT, since it gives high controlled quantum teleportation fidelity for highly mixed cases. These results  hold true for   different measures of multipartite entanglement. 
\nocite{*}
\section*{Acknowledgement}
K.G.P expresses his sincere gratitude to Dr.S. V. M. Satyanarayana for stimulating  discussions and acknowledges  financial support from the \textit{REDX}, Center for Artificial Intelligence, Indian Institute of Science Education and Research (IISER)  Kolkata, funded by \textit{Silicon Valley Community Foundation}, California, USA.

\bibliography{apssamp}
\end{document}